# Simple Method for Perfect Reconstruction QMF Filter Design

Fernando Martín-Rodríguez, Fernando Isasi-de-Vicente, Mónica Fernández-Barciela.
fmartin@tsc.uvigo.es, fisasi@tsc.uvigo.es, monica@tsc.uvigo.es.
atlanTTic research center for Telecommunication Technologies, University of Vigo,
Campus Lagoas Marcosende S/N, 36310 Vigo, Spain.

*Abstract*- **The purpose of this work is the design of FIR QMF (Quadrature Mirror Filters) filters of perfect reconstruction and odd number of coefficients (even order). By design, these filters will have linear phase and integer delay. These filter pairs have many applications in wavelet transforms and other multi-frequency decompositions. Perfect reconstruction filters have been studied from long time ago but we still found publications about them.**

*Keywords*- **Digital Filter Design, QMF (Quadrature Mirror Filters), FIR (Finite Impulse Response), Perfect Reconstruction.**

I. INTRODUCTION

In this paper we work on digital filter design. Concretely of the types: FIR (Finite Impulse Response) and QMF (Quadrature Mirror Filters) to yield perfect reconstruction of signals that are divided into sub-bands using them. FIR filters will be constructed with odd number of coefficients so that we get an even order in the z transform polynomial. By design, these filters will have linear phase and integer delay. These filter pairs have many applications in wavelet and other multi-frequency decompositions. Perfect reconstruction filters have been studied from long time ago but we still found publications about them [1,2].

The fundamental scheme for our study will be typical for two-channel QMF filters:

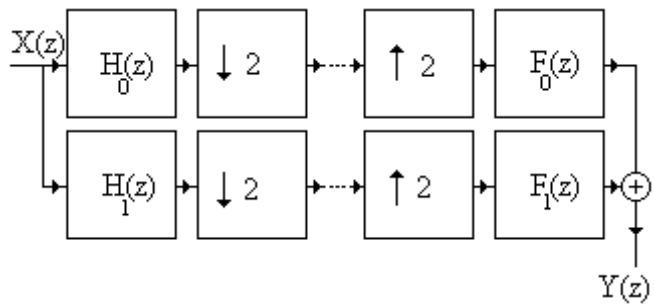

Fig. 1. General block diagram.

In the figure, we can see an analysis bank (H filters) and a synthesis one (F filters). We are assuming an ideal case in which encoding and transmission of information does not introduce any loss. This is the right model when we care only about filters. It can be easily demonstrated that the output $Y(z)$ can be computed as:

$$Y(z) = 1/2[H_0(z)F_0(z) + H_1(z)F_1(z)]X(z) \\ + 1/2[H_0(-z)F_0(z) + H_1(-z)F_1(z)]X(-z) \quad (1)$$

Reconstruction filters (F) are usually chosen as follows: $F_0(z) = H_1(-z)$ and $F_1(z) = -H_0(-z)$. This eliminates the component in $X(-z)$ and it can be assumed that the assembly becomes a single LTI system: $Y(z) = T(z)X(z)$, where $T(z)$ is given by:

$$T(z) = \frac{1}{2}[H_0(z)H_1(-z) - H_1(z)H_0(-z)] \quad (2)$$

Note that our goal is that $T(z)$ is a delay (perfect reconstruction). It should be noted that there are filters $H_0$ and $H_1$ that meet that condition but do not divide the signal into sub-bands. These filters are called "trivial solutions" and, although they are perfect reconstruction filters, they have no practical use. We wish the $H_0$ and $H_1$ filters resemble as much as possible an ideal low-pass and an ideal high-pass filter respectively.

There are only three possibilities to have perfect reconstruction and linear phase filters [3]:

a) Both filters are symmetrical and have an odd number of coefficients. The difference in their lengths is an odd multiple of 2.
b) One filter is symmetrical and the other antisymmetric, both have an even number of coefficients. Difference in lengths is an even multiple of 2 (possibly 0).
c) One filter has an odd number of coefficients and the other has an even number. Both filters have all zeros in the unit circle. Both can be symmetric or one is symmetric and the other antisymmetric. In this third case, only trivial solutions are obtained.

Case b) is known in the classical literature on the subject [4,5] and case a) is obviously the one that concerns us. A typical solution for case b is $H_1(z) = H_0(-z)$. Id EST: designing a (symmetric and odd length) low-pass $H_0$ and compute $H_1$ in this simple manner yields a good solution, but delay will not be an integer. We will explore a similar solution for case a: designing a low-pass (2n+1 length) and computing a high-pas (2n-1 length). We will see that this first high pass solution is unique and it is a "basis" for generating new (and perhaps better) solutions (of lengths: 2n+3, 2n+5…).

## II. Design Methods

### A. Initial Method

Let's define $P(z) = H_0(z)H_1(-z)$, according to equation 2, $T(z) = 1/2[P(z) - P(-z)]$. Trivially, in this subtraction the even powers will always cancel out while the odd ones are multiplied by 2. Our goal is that $T(z)$ is a delay, that is: we want all odd terms to be null (except the central one).

From this, we deduce the following design method: first design a low-pass filter $H_0$ with 2n+1 coefficients (in addition, it must be symmetric to have linear phase). We will assume that there is a high-pass couple $H_1$ of 2n-1 coefficients to be determined.

To determine them, we derive equations to cancel all the terms of the product $P(z)$ except the central one. Taking into account that all the polynomials involved are symmetric, we will have n linear equations with n unknowns. In fact, the expression for the coefficient of the i-th power of $z^{-1}$ in $P(z)$ is:

$$a_0(-1)^i b_i + a_1(-1)^{i-1} b_{i-1} + \cdots + a_i(-1)^0 b_0 \quad (3)$$

Where $a_i$'s are the coefficients of $H_0$ and the $b_i$'s are unknown coefficients for $H_1$. For odd values of i in the first half of $P(z)$ (i=1,3…2n-1), we can write equations ($b_i$'s as unknowns) equating expression (3) to zero, except for i=2n-1. This last equation corresponds to the central term in $P(z)$ and it will be forced to a value of 1.0 (in fact, it can be any non-zero value). Note that if all expression would be equated to zero, we would only get trivial solutions.

Given $H_0$, $H_1$ is unique except for a scale factor. That factor depends on the non-zero value chosen for the last equation (see above). Anyway, $H_1$ it is normalized to get unit gain in the pass-band.

If this method is implemented as is, results will not be good. The quality of $H_0$ will be that of the method used to design it; however, $H_1$ will be defective. The resulting $H_1$ filter has an adequate passband response but has a peak at w=0. In the next point, we will see how this situation can be corrected.

### B. Refinement

Refinement is based on looking for other valid $H_1$'s valid with more coefficients. It can be shown [6] that these other filters can be found from the "basic" $H_1$ using the following expression:

$$H_1'(z) = z^{-2m} H_1(z) + E(z) H_0(z) \quad (4)$$

Where $E(z)$ is any symmetric polynomial of order 4m-2 that has only even powers of z (m coefficients are distinct and not zero). Looking at equation (4), we can see that it adds to $H_1$ a term proportional to $H_0$. Id EST: that term is low-pass and can correct the defects of $H_1$ in the stop band.

Taking m=1, $E(z)$ depends on a single scalar and it is easy to adjust their value so that it can eliminate the peak at 0. In fact, choosing the right value to cause a transmission zero at w=0 (z=1) yields good results (section III).

Testing with filters of many coefficients is not enough to ensure a zero at w=0. In these cases, a correction with m>1 is required. For example, for m=2, $E(z)H_0(z)$ can be decomposed as sum of two terms: $e_0(1+z^{-6})H_0(z)$ and $e_2(z^{-2}+z^{-4})H_0(z)$. In this case, we can set up and solve a system of 2 linear equations to ensure two zeros in the stop band. In general, the system will be of m equations and we will have m transmission zeros in the stop band.

## III. Results

We started by testing with a very simple $H_0$ design: a symmetrical FIR filter created from a delayed ideal low-pass that is windowed with different kind of windows: rectangular, hamming, gauss… This is a common method of FIR filter deign by windowing [7].

Nevertheless, this method does not produce sharp transitions. We changed to a new model where the symmetric filter is created through the subtraction of two squared **sinc** functions $(sinc(x) = \sin(\pi x)/(\pi x))$. In the frequency domain this is equivalent to a subtraction of two slope triangles that yields a trapezoid (adjusting constants adequately this can give a very good filter). Equations for his design are:

$$C0 = \frac{w_{s0}^2}{2\pi(w_{s0}-w_{p0})}$$
$$B0 = \frac{w_{p0}^2}{2\pi(w_{s0}-w_{p0})}$$
$$h[n] = C0\, \text{sinc}^2\left(\frac{w_{s0}\, n}{2\pi}\right) - B0\, \text{sinc}^2\left(\frac{w_{p0}\, n}{2\pi}\right) \quad (5)$$

Where $w_{p0}$ will be the end of the pass-band and $w_{s0}$ the start of the stop-band.

For n=10 (21 coefficients in the low-pass) and m=1 the result of Figure 2 was obtained (rectangular window). You can see that the fall is strong but there is rather rippled in stop band. Computing the MSE (Mean Square Error of the frequency response comparing with the ideal case), we obtain 0.047 for the low pass and 0.017 for the high pass. In logarithmic units: $-10 \log_{10}(MSE)$, results are 13.24 dB for low pass and 17.77 dB for the high pass.

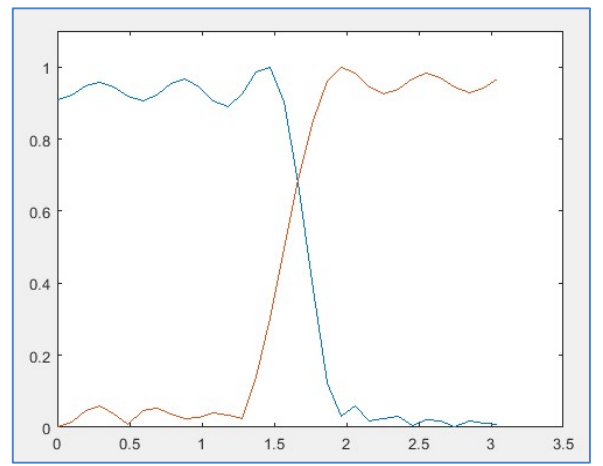

Fig. 2. Result for rectangular window.

If we change to Hamming window, the ripple is softened but we pay it, having a weaker fall (see figure 3 where n and m are maintained). MSE now is 13.80 dB for low-pass and 16.04 dB for high-pass. The Gaussian window gives us an intermediate point (see figure 4 for equal values of n and m), with MSE of 13.56 dB for low-pass and 17.58 dB for high-pass.

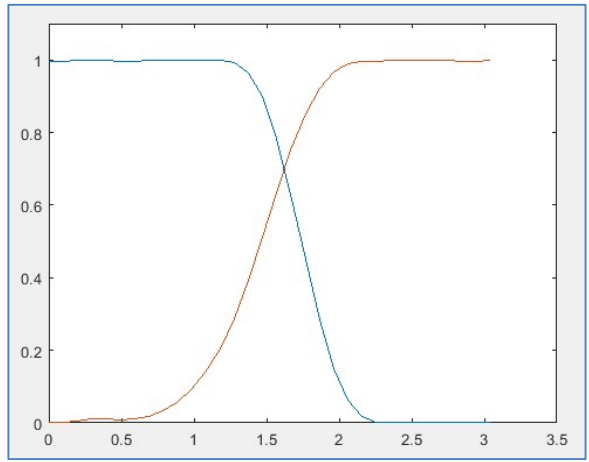

Fig. 3. Result for Hamming window.

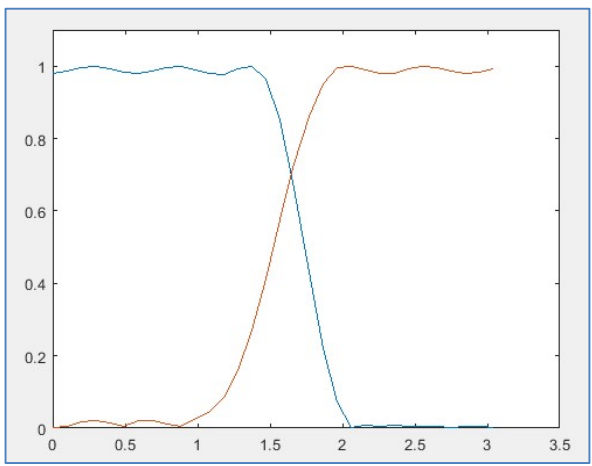

Fig. 4. Result for gaussian window.

Finally (figure 5), we see the results with Kaiser window (n = 20, 41 coefficients in the low pass, m = 2). MSE now goes to 12.56 dB for low-pass and 17.70 dB for high-pass.

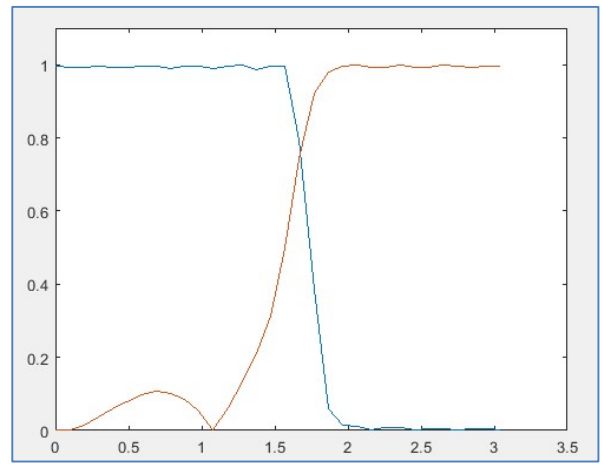

Fig. 5. Result for Kaiser window.

## IV. CONCLUSIONS AND FUTURE LINES

We have designed filters that meet the specifications with a very simple method. The behavior of them is acceptable and the practical applications are many, mainly: wavelet transform computation and other multi-frequency decompositions for coding or processing.